\documentclass{webofc}
\usepackage[varg]{txfonts}   % Web of Conferences font
\usepackage{amsfonts}
\usepackage{amsmath,epsfig,bm}
\usepackage{graphicx}
\usepackage{bm}
\usepackage{color}

\usepackage{tabularx}
\usepackage{graphicx}
\usepackage{caption}

\newcommand{\be}{\begin{equation}}
\newcommand{\ee}{\end{equation}}
\newcommand{\ba}{\begin{eqnarray}}
\newcommand{\ea}{\end{eqnarray}}
\newcommand{\bd}{\begin{displaymath}}
\newcommand{\ed}{\end{displaymath}}

\begin{document}
\title{Anomalous kaon correlations measured in Pb-Pb collisions at the LHC as evidence for the melting and refreezing of the QCD vacuum}

\author{\firstname{Joseph} \lastname{Kapusta}\inst{1}\fnsep\thanks{\email{kapusta@umn.edu}} \and
        \firstname{Scott} \lastname{Pratt}\inst{2}\fnsep\thanks{\email{prattsc@msu.edu}} \and
        \firstname{Mayank} \lastname{Singh}\inst{1,3}\fnsep\thanks{\email{mayank.singh@vanderbilt.edu}}
        % etc.
}

\institute{School of Physics \& Astronomy, University of Minnesota, Minneapolis, MN 55455, USA 
\and
           Department of Physics and Astronomy and Facility for Rare Isotope Beams, Michigan State University, East Lansing, MI 48824, USA 
\and
           Department of Physics and Astronomy, Vanderbilt University, Nashville, TN 37240, USA
          }

\abstract{%
Measurements of the dynamical correlations between neutral and charged kaons in central Pb-Pb collisions at $\sqrt{s_{NN}} = 2.76$ TeV by the ALICE Collaboration display anomalous behavior relative to conventional heavy-ion collision simulators.  We consider other conventional statistical models, none of which can reproduce the magnitude and centrality dependence of the correlations.  The data can be reproduced by coherent emission from domains which grow in number and volume with increasing centrality.  We study the dynamical evolution of the strange quark condensate and show that the energy released during the expansion and cooling of the system may be sufficient to explain the anomaly.
}
\maketitle
\section{Introduction}
\label{intro}

The search for experimental evidence of coherent fields related to the QCD phase transition has attracted much attention in relativistic heavy-ion collision physics.  
Recently the ALICE Collaboration published results for fluctuations of charged versus neutral kaons for Pb+Pb collisions at $\sqrt{s_{NN}} = 2.76$ TeV \cite{ALICEnu}. The magnitude of the measurement well exceeded expectations from typical effects such as charge conservation and Bose-Einstein symmetrization 
\cite{Kapusta:2022ovq}, and could not be reproduced by conventional event generators \cite{Nayak:2019qzd}. The observable analyzed by ALICE is known as $\nu$-dynamic \cite{Gavin:2001uk}
\begin{equation}
\nu_{\rm dyn}=
  \frac{\langle N_\pm(N_\pm-1)\rangle}{\langle N_\pm\rangle^2}
+ \frac{\langle N_S(N_S-1)\rangle}{\langle N_S\rangle^2}
-2\frac{\langle N_SN_\pm\rangle}{\langle N_S\rangle\langle N_\pm\rangle}.
\end{equation}
Here $N_\pm$ is the number of observed charged kaons and $N_S$ is the number of observed neutral kaons, which in this case are only $K_S$ mesons because ALICE cannot measure $K_L$ mesons. 

If $\nu_{\rm dyn} > 0$ detection of one particle biases the next particle to be of the same type. It is the opposite for $\nu_{\rm dyn} < 0$.  It is considered a relatively robust observable.  To scale the result in such a way that cancels the usual inverse scaling of correlations with multiplicity, ALICE presents $\nu_{\rm dyn}$ scaled by the factor $\alpha=(\langle N_\pm\rangle+\langle N_S\rangle)/\langle N_\pm\rangle\langle N_S\rangle$. 
For the most central collisions, ALICE's measurement of $\nu_{\rm dyn}/\alpha \approx 0.2$ is about five times the expected value. Furthermore, the additional charged, or neutral, kaon production appears to be spread uniformly in rapidity. 
Here we report on the new physics that might be underlying these experimental results \cite{Kapusta:2022ovq,DIC}.

\section{Modeling the Data}
\label{sec-1}

The data taken by ALICE is shown in Fig. 1.  
\begin{figure}[h]
\centering
\sidecaption
\includegraphics[width=6.5cm,clip]{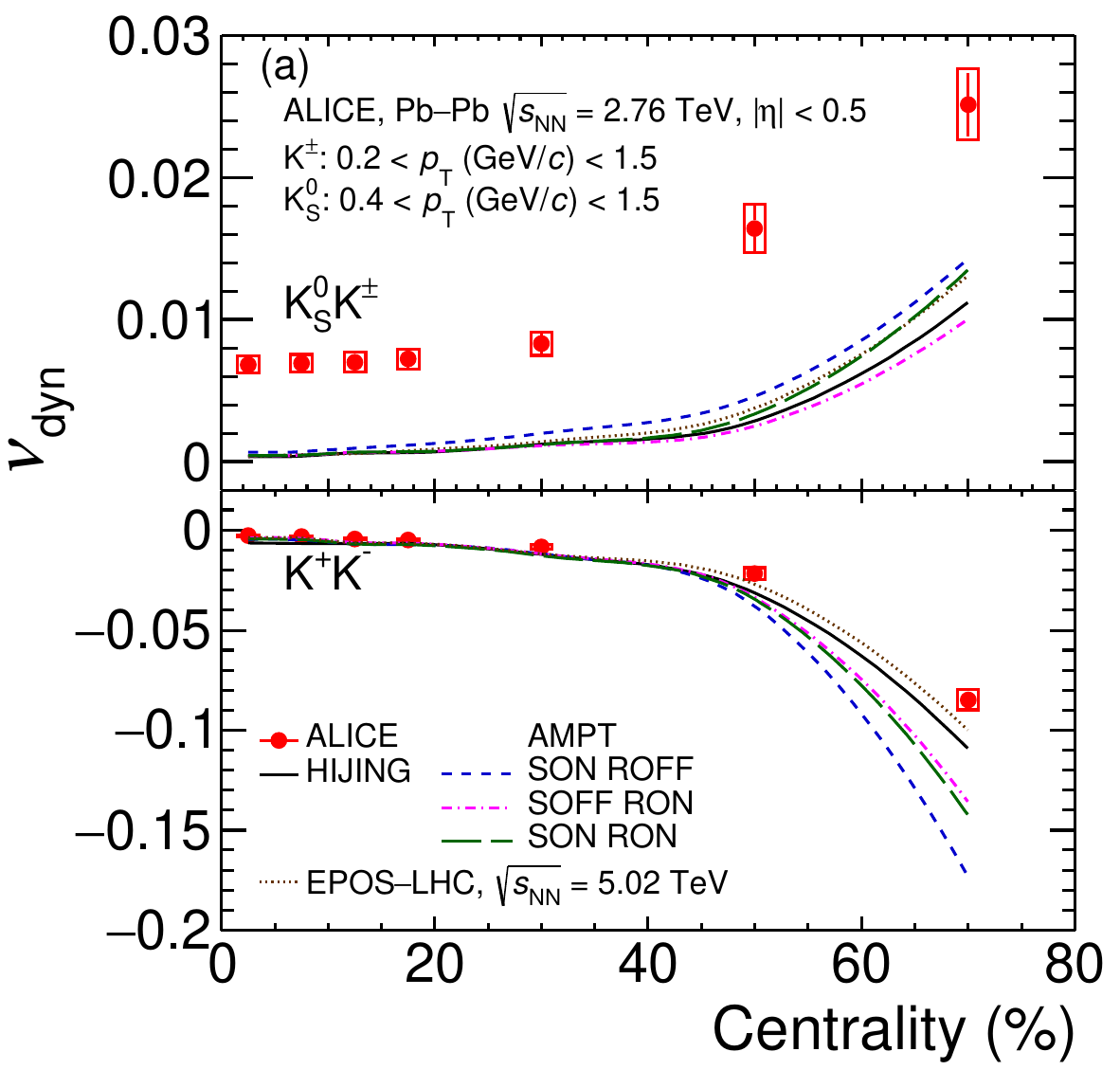}
\caption{The observable $\nu_{\rm dyn}$ for charged and neutral-short kaons as measured by the ALICE Collaboration \cite{ALICEnu}.  The data were compared to the event generators HIJING, EPOS-LHC, and three versions of AMPT.}
\label{fig-1} 
\end{figure}

In order to model this data, suppose we have multiple independent domains of condensates which each have a flat neutral kaon fraction probability distribution $P(f) = 1$ where $f$ is the fraction of kaons which are electrically neutral.  This is the case for strange DCC with three flavors \cite{JJ} or where a fraction of kaons occupy a single quantum state \cite{Kapusta:2022ovq}.  If the number of domains $N_d$ is greater than 2 or 3 then \cite{Gavin:2001uk} 
\begin{equation}
\nu_{\rm dyn} = 4 \beta_K \left( \frac{\beta_K}{3N_d} - \frac{1}{N_K^{tot}} \right)        
\end{equation}
where $\beta_K$ is the fraction of all kaons that come from condensate domains.  This relation is derived by folding the distributions of kaons from condensates and thermal/random sources. For multiple condensate sources, $P(f)$ approaches a Gaussian by the Central Limit Theorem.  Assuming reasonable scaling results in 
$\beta_K = (\epsilon_\zeta V_d)/(m_K N_K^{tot})$ where $\epsilon_\zeta$ is the energy density available from condensation and where $V_d$ is the sum total volume of all condensates.  This results in the two parameter formula
\begin{equation}
\frac{\nu_{\rm dyn}}{\alpha} = \frac{2}{3} b \left( \frac{\tau_{av}}{10 \tau_0} \right) \left[ \frac{b}{3a} \left( \frac{\tau_{av}}{10 \tau_0} \right)
 - 1 \right] .
\end{equation}
We obtain the fireball lifetime $\tau_{av}/\tau_0$ as a function of centrality from realistic hydrodynamic simulations of heavy-ion collisions using MUSIC \cite{Schenke:2010nt} with IP Glasma initial conditions \cite{Schenke:2012wb}.  The results are displayed in Fig. 2 and Table 1.  We fit only the five most central bins.  The reason is that the next bin has less than two domains within the acceptance, which violates the assumptions of the model.  Further study is needed for the larger impact parameters.

\begin{figure}
  \begin{minipage}[b]{.5\linewidth}
    \centering
    \includegraphics[width=1.0\linewidth]{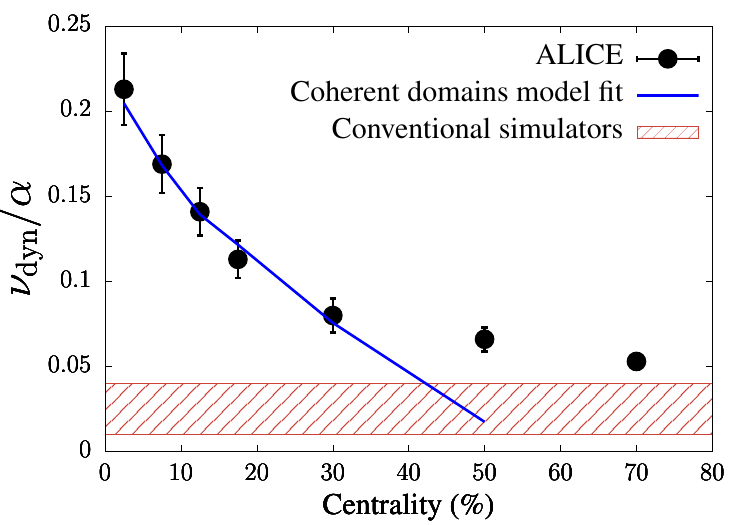}
    \captionof{figure}{Two parameter fit to the five highest multiplicity bins.  Data is from \cite{ALICEnu}.}% \caption{Figure caption}
\label{fig-2}
  \end{minipage}\hfill
  \begin{minipage}[b]{.45\linewidth}
    \centering
    \begin{tabular}{|c|c|c|c|c|}
\hline
Centrality & $N_d$ & $V_d$(fm$^3$) & $\beta_K$ \\
\hline
%\hline
%%%%%%%%%%%%%%%%%%%%%%%%%%%%%%%%%%%%%%%%%%%%%%%%
0-5 \% & 9.32 & 1120 & 0.302 \\
\hline
%%%%%%%%%%%%%%%%%%%%%%%%%%%%%%%%%%%%%%%%%%%%%%%%
5-10 \% & 7.29 & 821 & 0.283 \\
\hline
%%%%%%%%%%%%%%%%%%%%%%%%%%%%%%%%%%%%%%%%%%%%%%%%
10-15 \% & 6.02 & 640 & 0.267 \\
\hline
%%%%%%%%%%%%%%%%%%%%%%%%%%%%%%%%%%%%%%%%%%%%%%%%
15-20 \% & 4.67 & 476 & 0.256 \\
\hline
%%%%%%%%%%%%%%%%%%%%%%%%%%%%%%%%%%%%%%%%%%%%%%%%
20-40 \% & 2.88 & 258 & 0.225 \\
\hline
%%%%%%%%%%%%%%%%%%%%%%%%%%%%%%%%%%%%%%%%%%%%%%%%
40-60 \% & 1.20 & 82 & 0.172 \\
\hline
%%%%%%%%%%%%%%%%%%%%%%%%%%%%%%%%%%%%%%%%%%%%%%%%
\end{tabular}
    \captionof{table}{The number of sources or domains $N_d$ and the total volume $V_d$ occupied by them from the two parameter fit.  This assumes $\epsilon_{\zeta} = 25$ MeV/fm$^3$ to infer the volume.}
\label{tabl-1}
  \end{minipage}
\end{figure}

\section{DCC versus DIC}

It is always assumed that $\langle \bar{u}u \rangle = \langle \bar{d}d \rangle$.  What if their relative magnitudes fluctuated at finite temperature?  This means fluctuations between an isosinglet $\langle \bar{u}u \rangle + \langle \bar{d}d \rangle$ and an isotriplet $\langle \bar{u}u \rangle - \langle \bar{d}d \rangle$.  If the domain happened to be totally $\langle \bar{u}u \rangle$ then, when it loses energy due to cooling, combination with strange quarks and anti-quarks results in charged kaons.  If the domain happened to be totally $\langle \bar{d}d \rangle$ then combination with strange quarks and anti-quarks results in neutral kaons.  If the distribution in the relative proportion of the two condensates was flat then we essentially recover the previous phenomenology.

\begin{figure}
\centering
\begin{minipage}{.5\textwidth}
  \centering
  \includegraphics[width=0.99\linewidth]{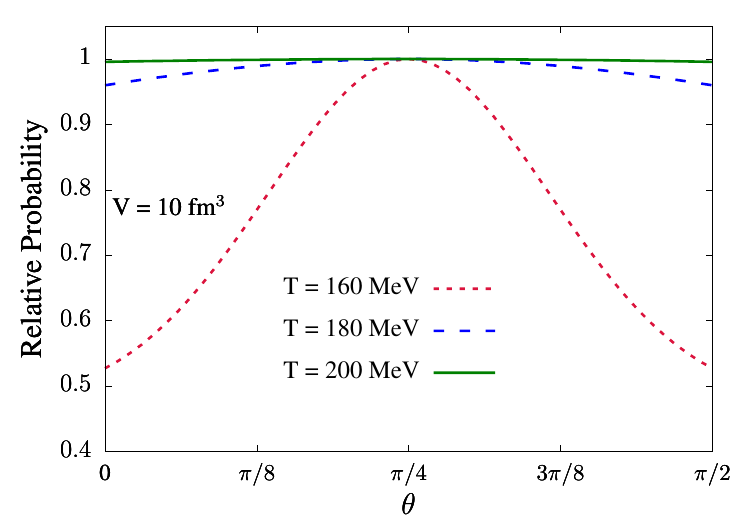}
\end{minipage}%
\begin{minipage}{.5\textwidth}
  \centering
  \includegraphics[width=0.99\linewidth]{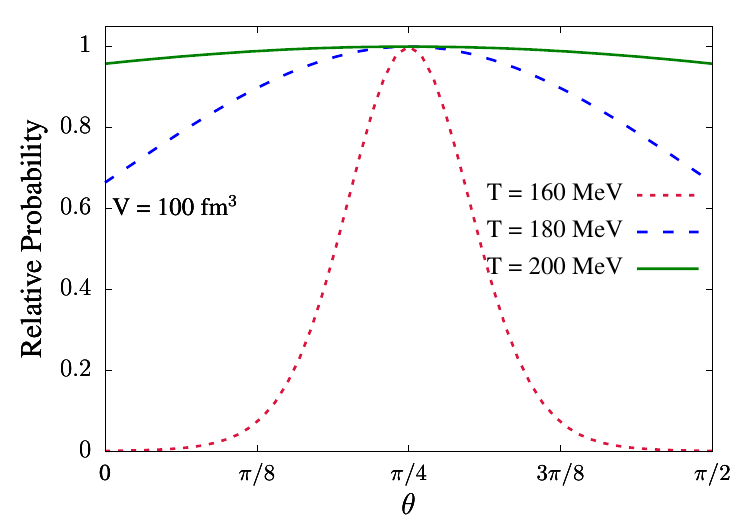}
 \end{minipage}
\caption{The relative probability $P(\theta)={\rm e}^{-V\Delta U/T}$ as a function of $\theta$ for several temperatures. Results are shown for two domain volumes, $V=10$ fm$^3$ (left) and $V=100$ fm$^3$ (right).}
\label{fig-3}
\end{figure}

To calculate the statistical distribution of the up and down quark condensates we utilize the 2+1 flavor linear sigma model as expounded in Ref. \cite{JJ}.  The condensates are related to the scalar fields by $\sigma_u = - \langle \bar{u}u \rangle/\sqrt{2}c'$, $\sigma_d = - \langle \bar{d}d \rangle/\sqrt{2}c'$, and $\zeta = - \langle \bar{s}s \rangle/\sqrt{2}c'$ where $c'$ is a constant.  We write $\sigma_u = \sigma \, \cos\theta$ and $\sigma_d = \sigma \, \sin\theta$ with $0 \le \theta \le \pi/2$, and take the temperature dependence of $\sigma$ and $\zeta$ from lattice calculations \cite{HotQCD1}.  The cost in free energy per unit volume is \cite{DIC}
\begin{equation}
\Delta U(T,\theta) = \frac{1}{2}\lambda \left[1-\sin^2(2\theta) \right]\sigma^4 
 + c(T) \left[1-\sin(2\theta)\right]\sigma^2\zeta 
 + f_\pi m_\pi^2 \left[ 1-\frac{\cos\theta+\sin\theta}{\sqrt{2}} \right]\sigma .
\end{equation}
The parameters and function $c(T)$ were determined in that paper.  For a domain of volume $V$ the relative probability for it to have the value $\theta$ relative to the equilibrium value $\pi/4$ is
\begin{equation}
P(\theta)={\rm e}^{-V\Delta U(\theta)/T} .
\end{equation}
Results for two representative domain volumes are shown in Fig. \ref{fig-3}.  For 10 fm$^3$ the distribution is essentially flat for $T \ge 180$ MeV, while for 100 fm$^3$ the distribution is essentially flat for $T \ge 190$ MeV.  The lattice calculations \cite{HotQCD1} show that $\sigma(T)$ has about 1/2 its vacuum value at $T \approx 160$ MeV and decreases to a negligible value at $T \approx 200$ MeV.  This means that a nontrivial amount of energy must be released, mainly in the form of light quarks and antiquarks whose flavor depends on the relative proportion of $\langle \bar{u}u \rangle$ and $\langle \bar{d}d \rangle$.  Of course a fully dynamical theory needs to be developed and implemented in order to make more precise predictions.

\newpage

\section{Conclusions}

ALICE has measured isospin correlations in the kaon sector which are anomalously large.  These measurements cannot be explained by any known means without invoking kaon condensation (least likely), Disoriented Chiral Condensates (less likely), or Disoriented Isospin Condensates (most likely).  DCC involve disorientation in the strange quark sector while DIC involve disorientation in the light quark sector.  It should also be acknowledged that quarks and anti-quarks are most likely strongly correlated already before chemical freezeout.  It would be illuminating to see similar measurements at $\sqrt{s_{NN}}= 5.02$ TeV Pb+Pb collisions at LHC and at  
$\sqrt{s_{NN}}= 200$ GeV Au+Au collisions at RHIC.  More differential measurement in rapidities and azimuthal angles are needed.  Can lattice QCD contribute?  Are we seeing the melting and refreezing of the QCD vacuum?\\

\section*{Acknowledgments}

The work of JK and MS was supported by the U.S. Department of Energy Grant No. DE-FG02-87ER40328, and the work of SP was supported by the U.S. Department of Energy Grant No. DE-FG02-03ER41259. MS is also supported by U.S. Department of Energy Grant No. DE-SC-0024347.

%
% BibTeX or Biber users please use (the style is already called in the class, ensure that the "woc.bst" style is in your local directory)
% \bibliography{name or your bibliography database}

\end{document}